\documentclass[twocolumn]{aastex6}

\shorttitle{SETI with Gaia}
\shortauthors{Zackrisson et al.}

\begin{document}


\title{SETI with Gaia: The observational signatures of nearly complete Dyson spheres}


\author{Erik Zackrisson\altaffilmark{1}$^*$, Andreas J.\ Korn\altaffilmark{1}, Ansgar Wehrhahn\altaffilmark{1}}
\affil{Department of Physics and Astronomy, Uppsala University, Box 515, SE-751 20 Uppsala, Sweden}
\author{Johannes Reiter\altaffilmark{2}}
\affil{University of Heidelberg, Germany}
\altaffiltext{*}{E-mail: erik.zackrisson@physics.uu.se}

\begin{abstract}
A star enshrouded in a Dyson sphere with high covering fraction may manifest itself as an optically subluminous object with a spectrophotometric distance estimate significantly in excess of its parallax distance. Using this criterion, the Gaia mission will in coming years allow for Dyson-sphere searches that are complementary to searches based on waste-heat signatures at infrared wavelengths. A limited search of this type is also possible at the current time, by combining Gaia parallax distances with spectrophotometric distances from ground-based surveys. Here, we discuss the merits and shortcomings of this technique and carry out a limited search for Dyson-sphere candidates in the sample of stars common to Gaia Data Release 1 and RAVE Data Release 5. We find that a small fraction of stars indeed display distance discrepancies of the type expected for nearly complete Dyson spheres. To shed light on the properties of objects in this outlier population, we present follow-up high-resolution spectroscopy for one of these stars, the late F-type dwarf TYC 6111-1162-1. The spectrophotometric distance of this object is about twice that derived from its Gaia parallax, and there is no detectable infrared excess. While our analysis largely confirms the stellar parameters and the spectrophotometric distance inferred by RAVE, a plausible explanation for the discrepant distance estimates of this object is that the astrometric solution has been compromised by an unseen binary companion, possibly a rather massive white dwarf ($\approx 1\ M_\odot$). This scenario can be further tested through upcoming Gaia data releases.   
\end{abstract}

\keywords{Extraterrestrial intelligence -- stars: distances -- methods: data analysis -- methods: spectroscopy -- dust, extinction -- stars: individual (TYC 6111-1162-1)}



\section{Introduction} 
\label{intro}
By constructing a Dyson sphere \citep{Dyson60} out of material from dismantled planets, extremely advanced civilizations could in principle tap into a significant fraction of the radiation power of their host star, thereby climbing the Kardashev scale towards type II status \citep{Kardashev64}. The Dyson sphere is typically not envisioned as a solid shell, but rather as a dense, spherical swarm or shroud of absorbing satellites, with each satellite absorbing a small fraction of the stellar radiation. Less ambitious astro-engineering designs, where sparse swarms, rings or single large shades are built around both stars and stellar remnants have also been considered \citep[e.g.][]{Suffern77,Wright14b,Osmanov16,Wright16}, and the time to build a small-scale megastructure of this type has been estimated at as little as $\sim 10^2$ yr \citep{Armstrong13}. Proposed uses for the collected energy include supercomputing, vessel propulsion, the powering of artificial habitats and long-range communication \citep[e.g.][]{Suffern77,Sandberg99,Arnold05,Armstrong13}. 

In the field of Searching for Extraterrestrial Intelligence (SETI), a number of searches for individual Dyson spheres have already been carried out, without unraveling any compelling evidence for such structures in the Milky Way \citep{Slysh85,Timofeev00,Jugaku04,Carrigan09}. Similar searches for civilizations engaging in galaxy-spanning or even intergalactic colonization powered by Dyson sphere technology have also failed to turn up any tell-tale signatures of this type of astro-engineering \citep{Annis99,Wright14a,Wright14b,Griffith15,Zackrisson15,Garrett15,Olson17}. Depending on the sizes and distribution of the obscuring satellites, these structures may give rise to conspicuous temporal variations in the apparent brightness of the affected star \citep[e.g.][]{Arnold05,Wright16,Osmanov16}. However, most searches so far have focused on the overall, long-term changes in the spectral energy distribution of the star that the satellites may induce.

Following the AGENT formalism of \citet{Wright14b}, the energy budget of a Dyson sphere can be described as:
\begin{equation}
\alpha + \epsilon = \gamma + \nu,
\label{AGENT_eq}
\end{equation}
where $\alpha$ represents the radiation energy collected by the Dyson sphere, $\epsilon$ the energy produced by other means, $\gamma$ the thermal waste heat and $\nu$ the energy lost through non-photonic means (e.g. neutrinos, gravitational waves). Under the assumption that the collected starlight overwhelms all other forms of energy generated and that non-thermal losses are negligible, equation~(\ref{AGENT_eq}) simplifies to $\alpha \approx \gamma$. In observational terms, even a minor fraction of collected starlight can then generate a very pronounced infrared excess in the spectral energy distribution of the star, as long as the Dyson sphere operates at a temperature in the 50--1000\,K range. Consequently, most Dyson-sphere searches have focused on detecting this infrared signature. However, in case where losses due to non-thermal photons are significant (i.e. where $\nu$ is non-negligible), where the energy is somehow stored rather than radiated away or where the waste heat radiation is anisotropic (and not emitted in our direction), the infrared signature may be weaker than usually assumed. Here, we outline a method to search for partial Dyson spheres with weak or even absent waste heat signatures.  

Since a star surrounded by a partial Dyson sphere with large (but non-unity) covering fraction is expected to appear unusually faint for its type at optical/near-infrared wavelengths, its spectrophotometric distance is expected to be overestimated. A trigonometric parallax measurement, as provided by Hipparcos or Gaia would, on the other hand, still be able to provide an accurate distance estimate. Hence, it should be possible single out tentative Dyson-sphere candidates by comparing parallax distances to their spectrophotometric counterparts. With future Gaia data releases, both distance estimates can be obtained from Gaia data, but even with Gaia Data Release (DR) 1 \citep{2016A&A...595A...1G} it is possible to carry out a search of this kind by combining the parallax distances provided by the Tycho-Gaia Astrometric Solution (TGAS) with spectrophotometric distances from ground-based surveys like the Radial Velocity Experiment (RAVE; \citealt{Kunder17}) or GALAH \citep{Martell17}. Candidates for Dyson spheres identified this way can also be tested for mid-/far-infrared excess through a comparison of predicted and observed fluxes at mid/far-IR wavelengths from infrared telescopes like WISE, IRAS, Spitzer, Akari and Herschel.

In section~\ref{discrepancy}, we describe the method and its limitations. In section~\ref{DR1} we apply this method to the $\approx 2\times 10^5$ stars in common between Gaia DR1 and RAVE DR5 in search of objects with signatures that match those expected for partial Dyson spheres with high covering fractions. As an example of the type of object that is identified by this method, we discuss the properties of TYC 6111-1162-1 in section~\ref{TYC6111}, and present follow-up observations to confirm the stellar parameters derived by RAVE. Section~\ref{discussion} discusses the prospects of applying the method to much larger data sets in the future. Section~\ref{summary} summarizes our findings.

\section{Discrepancies between spectrophotometric and parallax distances for partial Dyson spheres}
\label{discrepancy}
Through the use of stellar-atmosphere modeling and theoretical stellar isochrones in conjunction with observed photometric fluxes and spectroscopic data at optical/near-infrared wavelength, it is possible to accurately determine stellar parameters, dust-attenuation factors and spectrophotometric distances $D_\mathrm{spec}$, or conversely spectrophotometric parallaxes $\pi_\mathrm{spec}$, for various types of stars \citep[e.g][]{Binney14}. This procedure basically assumes that once the observed optical flux $f_\mathrm{opt}$ has been corrected for dust, the relation between the intrinsic optical luminosity $L_\mathrm{opt}$ and its spectrophotometric distance is:
\begin{equation}
F_\mathrm{opt} = \frac{L_\mathrm{opt}}{4\pi D_\mathrm{spec}^2},
\label{Fopt_eq1}
\end{equation} 
which implies 
\begin{equation}
D_\mathrm{spec} = \left( \frac{L_\mathrm{opt}}{4\pi F_\mathrm{opt}} \right)^{1/2}
\label{Dspec_eq1}
\end{equation}
However, consider a star enshrouded by a Dyson sphere which covers a fraction $f_\mathrm{cov}$ of the stellar surface area. If this obscuring structure blocks the emerging optical/near-IR light like a grey absorber (i.e. all wavelengths throughout this interval are equally affected), the only impact on the observed flux would be a general dimming, whereas the detailed shape of the spectrum at these wavelengths would be unaffected. The true distance would then be given by $D^\prime$:
\begin{equation}
D^\prime = \left( \frac{L_\mathrm{opt}(1-f_\mathrm{cov})}{4\pi F_\mathrm{opt}} \right)^{1/2}.
\label{Dtrue_eq1}
\end{equation}

Combining eqs.~\ref{Dspec_eq1} and ~\ref{Dtrue_eq1}, we get:
\begin{equation}
D^\prime  = D_\mathrm{spec} (1-f_\mathrm{cov})^{1/2}
\label{Dtrue_eq2}
\end{equation}

\begin{figure*}[t]
\plottwo{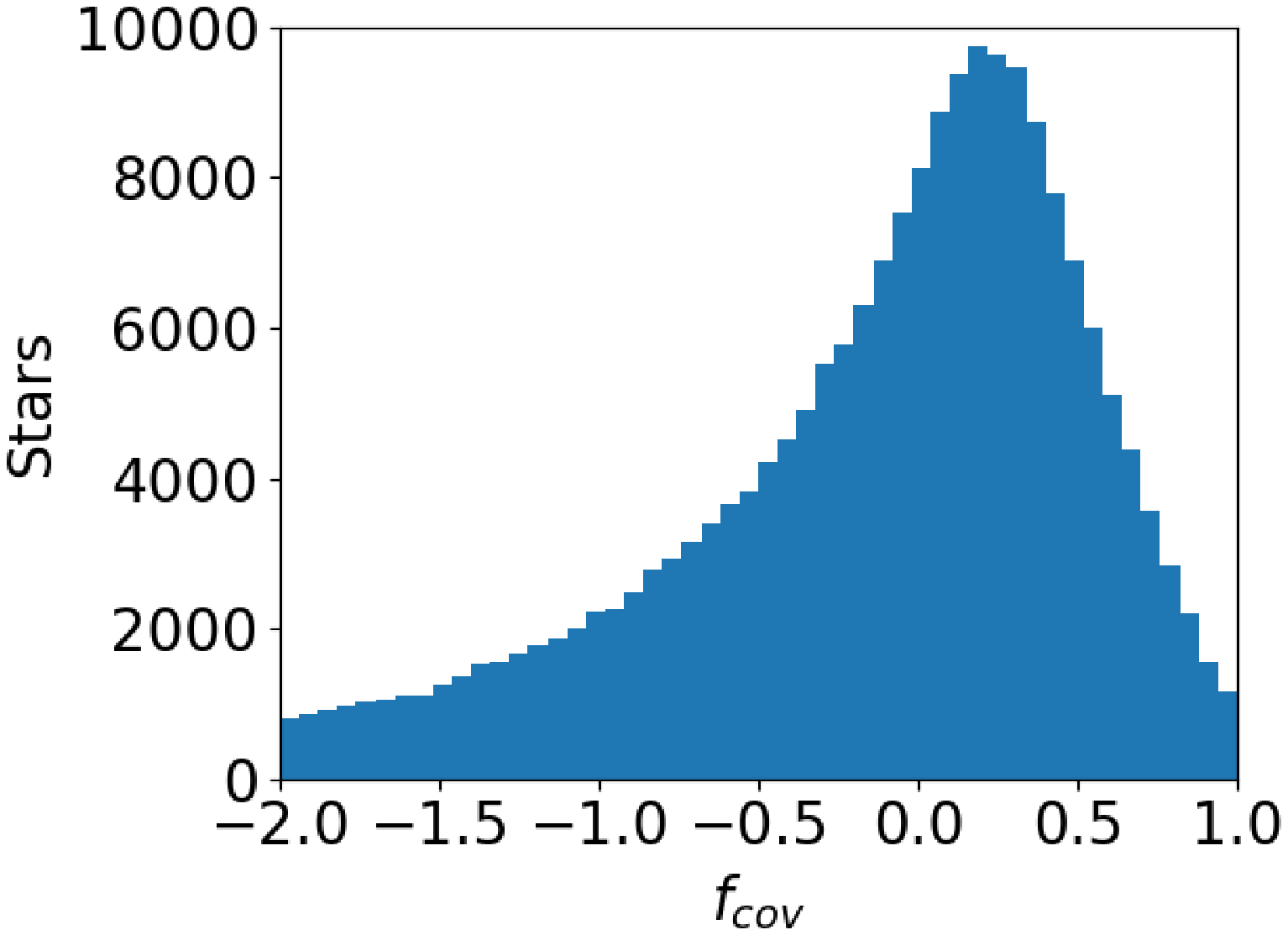}{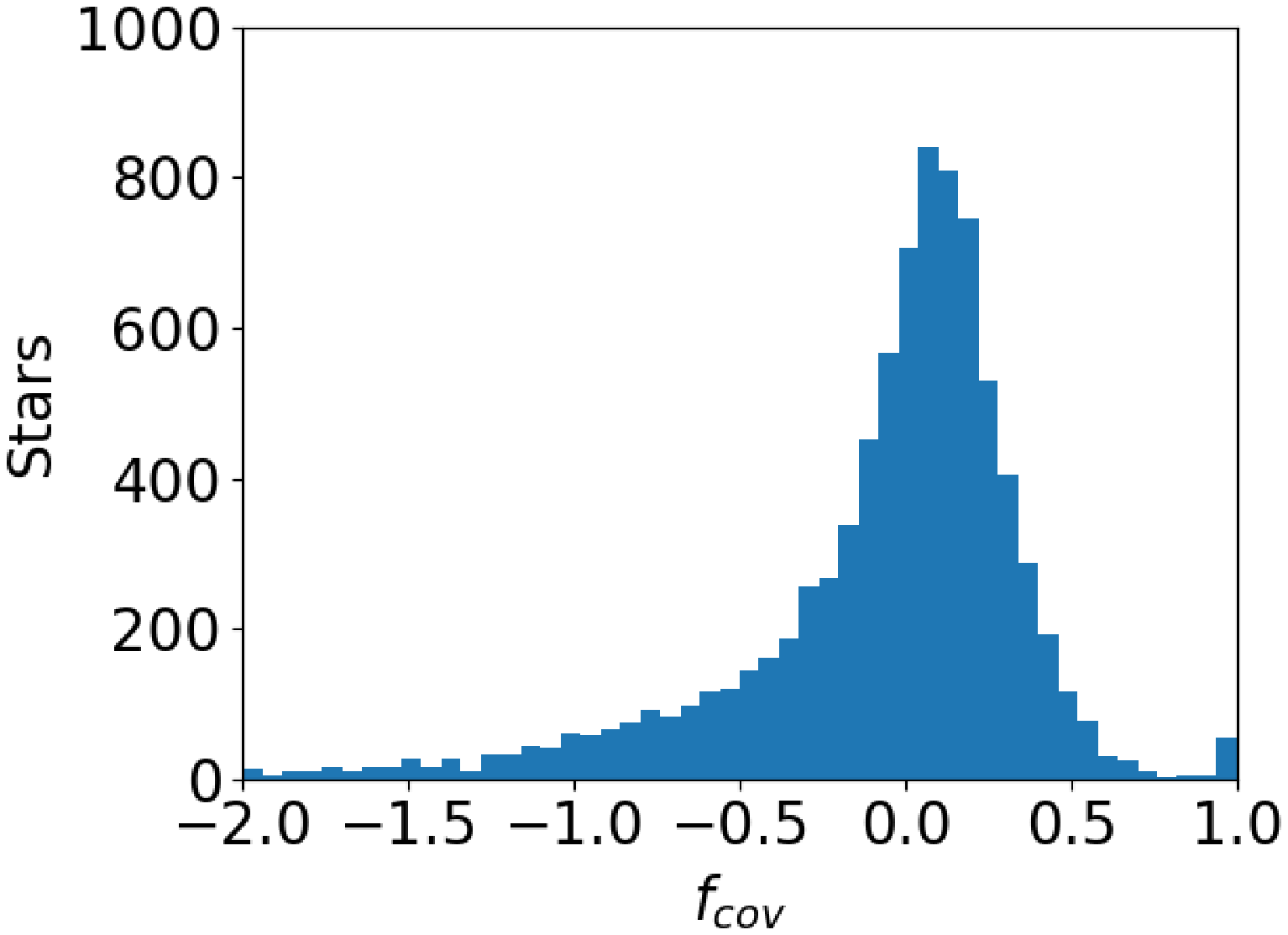}
\caption{Distribution of $f_\mathrm{cov}$ for {\bf a)} the whole Gaia DR1-RAVE DR5 sample; {\bf b)} the subset of the Gaia DR1-RAVE DR5 sample with the smallest parallax errors (trigonometric parallax error $\leq 10\%$, spectroscopic parallax error $\leq 20\%$).}
\label{fig_fcov_dist}
\end{figure*}

Hence, for a Dyson sphere which obscures a fraction $1>f_\mathrm{cov}>0$ of the intrinsic optical radiation, the true distance $D^\prime$ would be smaller than that inferred by standard spectrophotometric distance-estimation techniques. Even in this case, a measurement of the trigonometric parallax distance $D_\mathrm{trig}$ should still be able to retrieve the true distance, since this technique is not directly affected by the brightness of the star. If we adopt $D_\mathrm{trig}\approx D^\prime$ and use the relation between trigonometric parallax $\pi_\mathrm{trig}$ and distance $D$ ($\pi=1/D$, as long as the fractional parallax error is $\leq 0.2$, \citealt{Bailer-Jones2015}), then we can express $f_\mathrm{cov}$ as:
\begin{equation}
f_\mathrm{cov}  =  1-\left( \frac{D_\mathrm{trig}}{D_\mathrm{spec}} \right)^2 = 1-\left( \frac{\pi_\mathrm{spec}}{\pi_\mathrm{trig}} \right)^2
\label{fcov_eq}
\end{equation}

However, the square-root dependence in eq.~(\ref{Dtrue_eq2}) makes the expected difference between the two distances very small unless the star is close to being completely enshrouded, i.e.\ $f_\mathrm{cov}$ is close to unity. For example, $f_\mathrm{cov}\approx 0.1$ would correspond to a distance discrepancy of just $\approx 5\%$, which is undetectably small. In fact, it takes $f_\mathrm{cov}>0.75$ to produce a factor of $>2$ discrepancy, which may be more in the range of what one can hope to achieve given current errorbars.

In practice, to single out $f_\mathrm{cov}>0.75$ candidates from a large sample of objects requires that distances or parallaxes can be measured with relative errors of no more than $\approx 10$--$20\%$. This level of parallax precision is routinely achieved with space-based missions like Hipparcos or Gaia, and the Gaia mission is eventually expected to provide this level of parallax precision for $\sim 10^8$ main sequence stars \citep{Luri14}. The precision of spectrophotometric parallaxes is typically somewhat lower, with relative errors at the $\approx 20$--$30\%$ level \citep{Binney14} in the case of the RAVE survey, although the subset of stars with the very best estimates are still likely to be useful. 

In cases where a high-$f_\mathrm{cov}$ candidate is identified, the spectrophotometric fit also allows for a prediction of the intrinsic stellar flux at infrared wavelengths that can be directly compared to observed fluxes with mid/far-infrared survey data in search of any potential infrared excess. 

It should be noted that the covering fraction $f_\mathrm{cov}$ used in the above formalism does not need to represent the global covering fraction of the star, but simply the obscured fraction of the face of the star facing the observer. Hence, the principle not only applies to isotropic Dyson spheres, but also different types of stellar engines \citep[e.g.][]{Badescu00} and large transiting megastructures \citep{Wright16} that provide anisotropic coverage of the star. However, stellar engines that reflect some fraction of the light back at the host star -- like the Shkadov thruster \citep{Shkadov87,Forgan13} -- would likely affect its thermal equilibrium and may prevent stellar parameters from being accurately retrieved using standard spectral-analysis techniques.

\section{A search for partial Dyson-sphere candidates in Gaia data release 1}
\label{DR1}
For this pilot project, we have combined Gaia DR1 with the RAVE DR5 dataset. Gaia DR1 contains 5-parameter astrometric data for 2 million objects, whereas RAVE DR5\footnote{Stellar ages and masses are not provided in RAVE DR5, therefore these quantities are extracted from RAVE DR4} has spectroscopic data for over $4.8\times 10^5$ stars. Of these, about $2.3 \times 10^5$ objects feature in both catalogs. 

If one attempts to infer $f_\mathrm{cov}$ using eq.~(\ref{fcov_eq}) for the full data set, one arrives at the $f_\mathrm{cov}$ distribution shown in the left panel of Figure~\ref{fig_fcov_dist}, with extremely large numbers of stars at both very high and very low covering fractions. The reason for this is simply that many have distance errors much too large to allow meaningful constraints on $f_\mathrm{cov}$. By pruning the sample, and only keeping objects with parallax errors at $\leq 10\%$ for Gaia and  $\leq 20\%$ for RAVE (a good compromise between error and sample size, given the current data), one arrives at the much narrower distribution (featuring 8441 stars) shown in the right panel of Figure~\ref{fig_fcov_dist}. Once this cut has been applied, a small group of 75 stars (i.e. $\sim 1\%$ of the full sample) at $f_\mathrm{cov} > 0.9$ clearly stands out in the distribution. Under the assumption that parallax errors are Gaussian, one would naively expect only a fraction $\lesssim 0.04\%$ of the sample (i.e. $\lesssim 4$ objects) to spuriously appear at $f_\mathrm{cov}>0.9$. However, the probability distribution for the spectrophotometric distance modulus is in reality far more complicated and frequently features multiple peaks \citep[e.g.][]{Binney14}. All types of stars are moreover not equally well-described by current stellar-evolution tracks or stellar-atmosphere models, which makes it extremely difficult to reliably estimate the expected number of spurious outliers in the $f_\mathrm{cov}$ distribution. In the  following, we therefore take a closer look at the stars that end up in this high-$f_\mathrm{cov}$ tail in an attempt to further constrain their nature. 
\begin{figure}[t]
\plotone{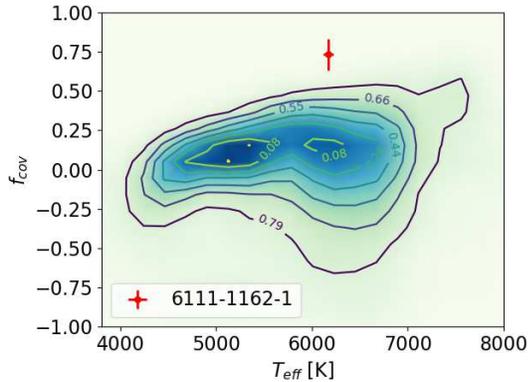}
\caption{Density of $f_\mathrm{cov}$ as a function of $T_\mathrm{eff}$ for all stars after the cut. The position of TYC 6111-1162-1 is indicated by a red cross. The numbers show the included fraction of stars for each contour.} 
\label{fig_fcov_Teff}
\end{figure}

Since the most interesting outliers would be those that belong to a spectral type for which current models are generally expected to produce highly reliable results, we are mainly interested in main-sequence FGK dwarfs. However, the vast majority of objects in the conspicuous $f_\mathrm{cov}>0.9$ peak in the right panel of Figure~\ref{fig_fcov_dist} have inferred values for the surface gravity $\log g < 2$ ($g$ in cgs units) and are consequently classified as low-gravity giants. Since it has been shown that the RAVE spectrophotometric distances systematically tend to then overestimate the distances to such stars \citep{Binney14}, albeit not to the level seen here, we will not consider these further. As an additional consistency check of the remaining outliers, we check the best-fitting mass, age and metallicity against the PARSEC \citep{2012MNRAS.427..127B} isochrones on which the RAVE fits were based. Occasionally, these stellar parameters yield nonsensical results when compared to the isochrones (e.g. a mass too high for the isochrone), which suggests some sort of glitch in the fitting. Once these are also removed, we are left with 6 outliers at $f_\mathrm{cov}>0.7$ out of 8365 stars. 

\begin{figure}[t]
\plotone{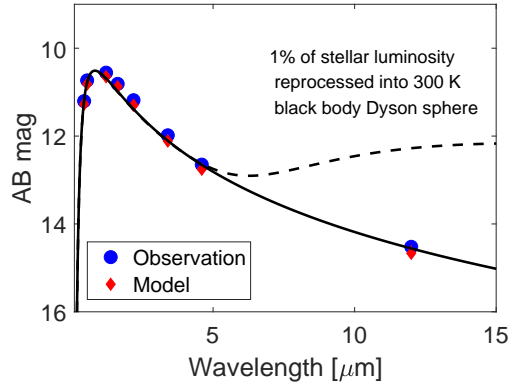}
\caption{Comparison of observed and predicted broadband optical-to-IR fluxes for TYC-6111-1162-1, based on the stellar parameters and spectrophotometric parallax presented in RAVE DR5. Blue circles indicate the observed optical (BV), near-IR (JHKs) and mid-IR (WISE W1, W2, W3) fluxes (expressed in AB magnitudes, with error bars smaller than the symbols used) and red diamonds the corresponding fluxes predicted by the PARSEC 1.1 isochrones given the best-fitting model (stellar parameters, interstellar extinction and spectrophotometric parallax) presented for this star in RAVE DR5. The black solid curve is a black body fit based on the bolometric luminosity and effective temperature of this stellar model. As seen, there is reasonable agreement between observed and modelled fluxes at all wavelengths plotted, and no indication of any significant IR excess. The dashed curve illustrates the strong IR excess expected even in a case where just 1\% of the bolometric stellar luminosity would be reprocessed into black body radiation by a Dyson sphere operating at 300 K.}
\label{SED_fig}
\end{figure}

To investigate these stars further, using additional observations from APASS (optical), 2MASS (near-IR) and WISE (mid-IR), we compare the observed Spectral Energy Distribution (SED), to the theoretical predictions from the PARSEC isochrones given the best-fitting stellar parameter. Four stars show conspicuous inconsistencies between the two, either indicating an incorrect RAVE distance modulus or a problem with the data in the observed catalogs. The only two stars remaining are TYC 7169-1532-1 and TYC 6111-1162-1, neither of which show any obvious mid-IR excess above the predictions from the stellar modeling. As discussed by \citet{McMillan17}, discrepancies between spectrophotometric and astrometric parallaxes exhibit a significant effective-temperature dependence, likely reflecting systematic problems and degeneracies in the spectrophotometric fitting procedure. In Figure~\ref{fig_fcov_Teff} we plot the inferred $f_\mathrm{cov}$ for our sample as a function of $T_\mathrm{eff}$. With its $T_\mathrm{eff}$ of $\approx$6200\,K, TYC-6111-1162-1 lies in the effective-temperature regime for which distance agreements tend to be very good, whereas TYC 7169-1532-1 with $T_\mathrm{eff}$ of 7500\,K is found in a region where distance overestimates seem to be far more common. We therefore take TYC 6111-1162-1 to be the most interesting individual candidate for further investigation in the current sample. 

\section{Case study: TYC 6111-1162-1}
\label{TYC6111}
In Figure~\ref{SED_fig}, we compare the observed apparent flux of TYC 6111-1162-1 at optical ($BV$), near-IR ($JHKs$) and mid-IR (WISE $W1$, $W2$, $W3$) wavelengths to the corresponding fluxes provided by the PARSEC 1.1 isochrones given the best-fitting model (stellar parameters, interstellar extinction and spectrophotometric parallax) for this star in RAVE DR5. As seen, this model provides an adequate fit to the data throughout the whole plotted wavelength range. The observed flux shows no detectable mid-IR flux excess compared to this model -- in stark contrast to the expectation for waste heat radiation from a Dyson sphere \citep{Dyson60}, even in the case where just a tiny fraction of the stellar luminosity is captured/reprocessed into waste heat. We exemplify the latter argument by showing the mid-IR excess (dashed line in Figure~\ref{SED_fig}) in a case where 1\% of the bolometric luminosity of the star is re-radiated as black body radiation at 300 K. This scenario would raise the flux at 12$\mu$m (the WISE $W3$ band) by more than two magnitudes, in direct conflict with the data, and can therefore be ruled out. Hence, the RAVE DR5 stellar model and spectrophotometric parallax provides a decent fit to the observed SED of TYC 6111-1162-1, without any conspicuous anomalies.

\begin{figure*}[t]
\includegraphics[width=5.3cm,angle=90]{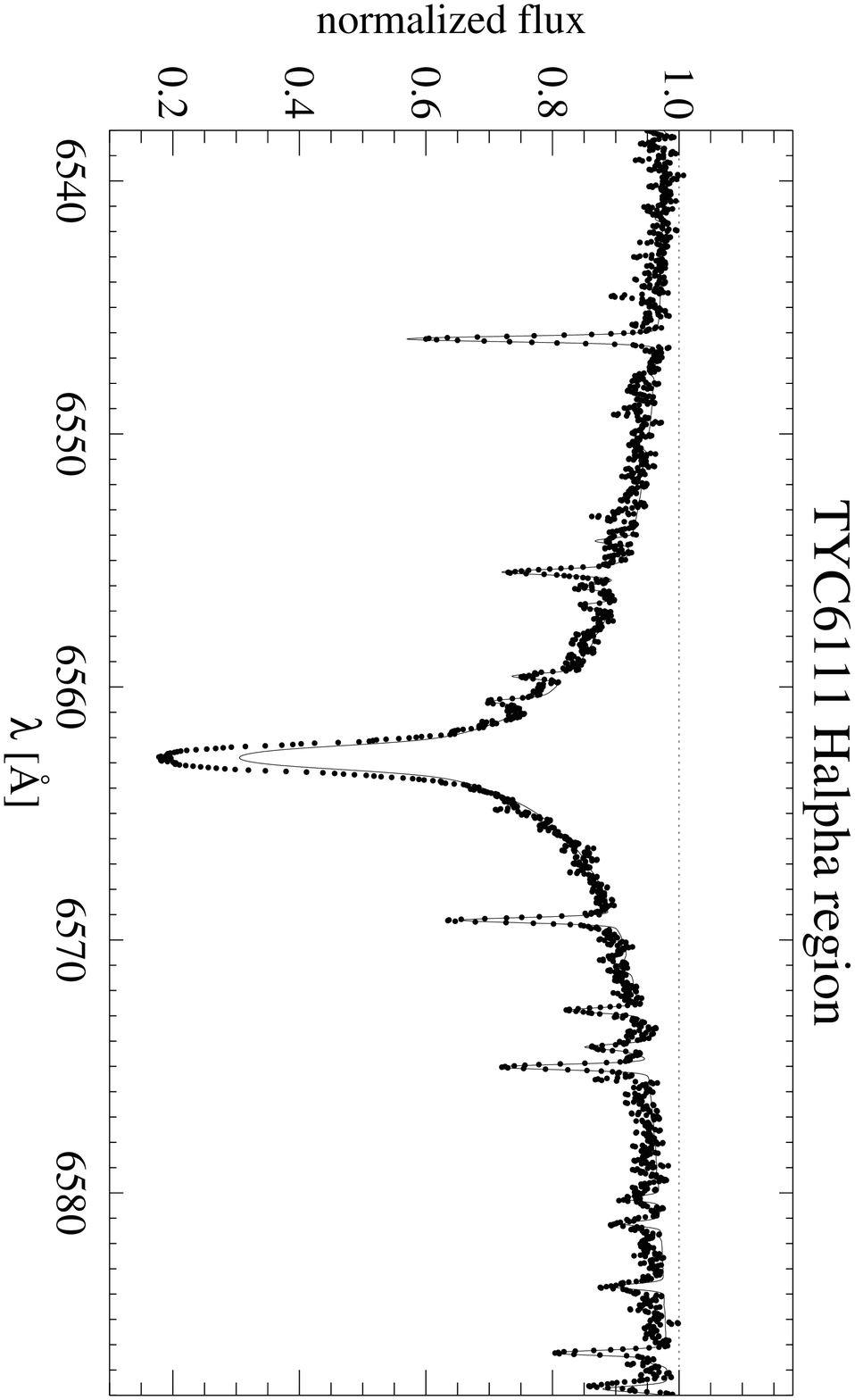}
\includegraphics[width=5.3cm,angle=90]{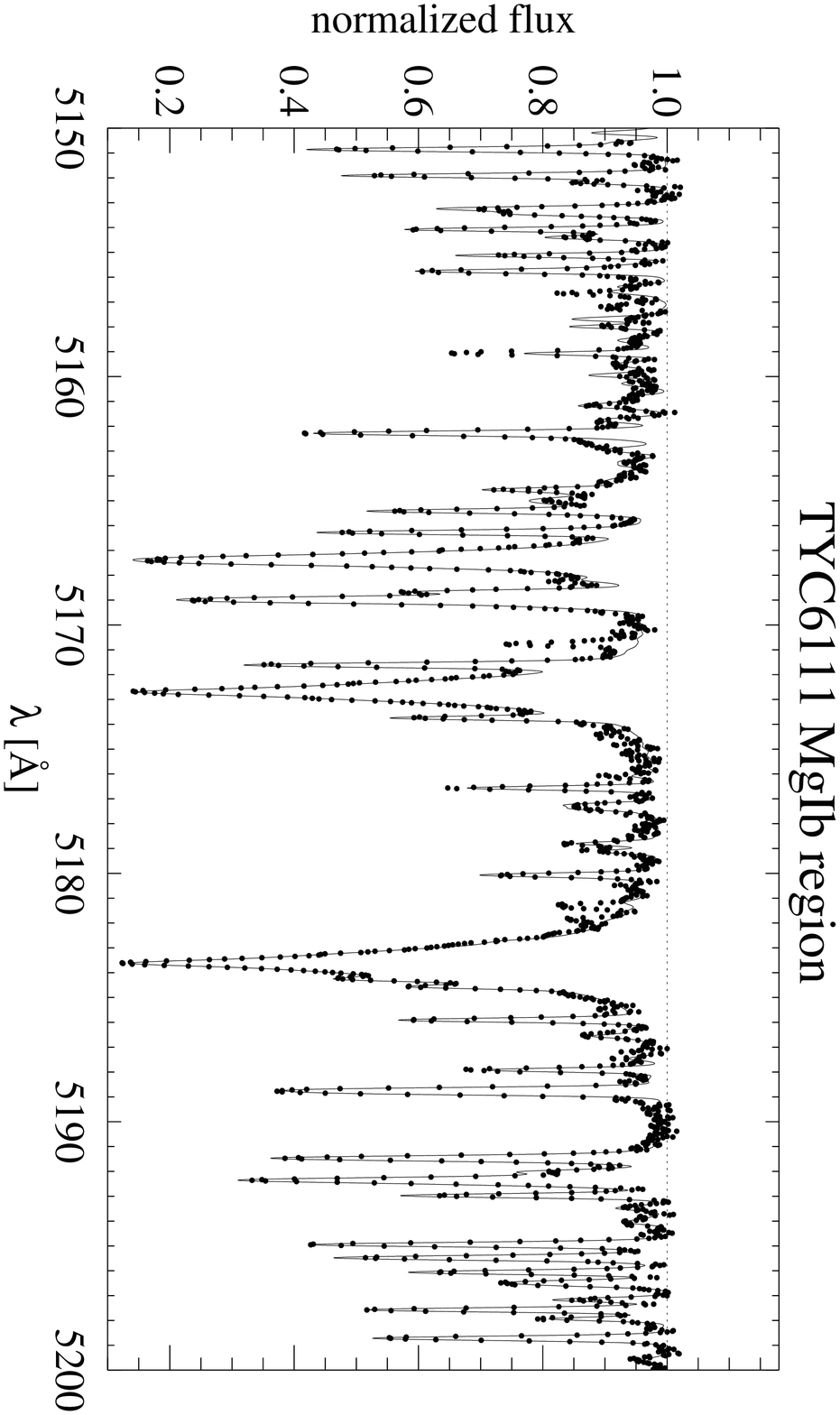}
\caption{Spectral regions of diagnostic power for the spectroscopic analysis of TYC 6111-1162-1: H$\alpha$, a sensitive indicator of the effective temperature ({\em left}) and the Mg I$b$ lines ({\em right\/}) that react sensitively to the surface gravity. Bullets mark the observations, the full-drawn line is the best synthetic fit to the data. The overall fit quality is very satisfactory (with the exception of the H$\alpha$ line core which is affected by the chromosphere and departures from LTE) indicative of a normal, metal-rich main-sequence star.}
\label{spectra}
\end{figure*}

To verify or refute the identified distance discrepancy between RAVE and Gaia, we obtained a higher-quality spectrum of this star. We took a 1h exposure of TYC 6111-1162-1 on February 9th, 2017 with the high-resolution spectrograph FIES \citep{Telting14} on the Nordic Optical Telescope (NOT). FIES provides continuous wavelength coverage from 370-830\,nm (to 910\,nm with some gaps) at a resolving power $R$\,=\,67,000. Data reduction was performed on-site using FIEStools \citep{Stempels17}. In the wavelength region used for the spectroscopic analysis (mostly $\lambda \geq 5000$\,\AA), the signal-to-noise ratio (SNR) is approximately 70 per pixel. 

We perform a line-by-line differential analysis relative to the Sun using tools very similar to those employed by \cite{Fuhrmann04}. In short, H$\alpha$ is used to constrain the effective temperature $T_\mathrm{eff}$ and the ionization equilibrium of neutral and ionized iron (Fe I/II) to constrain the surface gravity $\log g$ and the iron abundance [Fe/H]. The microturbulence $\xi$ is determined by requiring weak and strong iron lines to return the same abundance value. The model atmospheres used and the line-synthesis code assume LTE (local thermodynamic equilibrium). Only very small departure from LTE (at most a few 0.01\,dex in the log of the abundance) are to be expected in differential studies of stars this close to Solar parameters. This technique has proven very successful for local F- and G-type stars, with distances compatible with HIPPARCOS parallaxes to within 5\% (rms). Significantly larger offsets are a clear sign of stellar anomalies.

Starting from a longer list of iron lines that was shortened by an iterative 2$\sigma$-clipping procedure, 26 lines with line strengths (equivalent widths) between 28 and 103\,m\AA\ were used in the derivation of the final parameters. The resulting stellar parameters are: $T_\mathrm{eff}$ = 6150 $\pm$ 80\,K, $\log g$ = 4.24 $\pm$ 0.1, [Fe/H] = +0.32 $\pm$ 0.03, $\xi$ = 1.2 $\pm$ 0.1\,km/s (all uncertainties given as 1$\sigma$). As illustrated by Figure~\ref{spectra}, these stellar parameters describe the line spectra of selected diagnostic regions very well. Furthermore, the final $\log g$ (of importance for the spectrophotometric distance, see below) is fully supported by the pressure-sensitive wings of the Mg Ib lines, with the magnesium abundance determined from the weaker Mg I 4730 line. In summary, TYC 6111-1162-1 seems to be a perfectly normal, metal-rich main-sequence star of late F-type.

We can now compare our analysis with the results published by the RAVE consortium. \cite{Kunder17} list the DR5 results based on a SNR$\approx$45 spectrum as 6169$\pm$54\,K/4.54$\pm$0.11/+0.06$\pm$0.1 ($T_\mathrm{eff}$/$\log g$/[$m$/H]), while the data-driven RAVE-on analysis returns 6039$\pm$91\,K/4.29$\pm$0.13/+0.06$\pm$0.07. The agreement with both data set is good for $T_\mathrm{eff}$ and $\log g$, less so for the metallicity. Other metallicity indicators provided in RAVE DR5 (``Fe abund" and ``calib met") agree better at +0.18 and +0.16, respectively, and it seems justified to compare to one of these instead, as we are actually measuring iron. However, we find [Mg/Fe] = [Ca/Fe] = 0.00 $\pm$ 0.05 (based on Mg I 4730 and Ca I 6572) and thus see no reason why the Ca II triplet region used by RAVE should lead to a systematically underestimated metallicity. Our analysis based on better data may indicate that the parameter space of metal-rich stars is not (yet) well-calibrated in RAVE.

Having determined reliable spectroscopic stellar parameters, we can now use evolutionary tracks to determine the stellar mass. Interpolating among MIST stellar tracks \citep{Choi2016}, the star's mass is constrained to $M$ = 1.34 $\pm$ 0.05 ${\rm M}_\odot$ (RAVE DR4 lists 1.17 $\pm$ 0.1\,${\rm M}_\odot$ as the mass estimate for essentially the same set of stellar parameters as in DR5). Using $A_V$=0.10 following RAVE and interpolating in the tables of \cite{Alonso1995} to estimate an appropriate bolometric correction (BC$_V$=$-0.07$), we arrive at a spectrophotometric distance estimate of $D_{\rm spec}$ = 238 $^{+48}_{-34}$\,pc ($\pi_{\rm spec}$ = 4.2 $\pm$ 0.7\,mas). In RAVE DR5, the spectrophotometric parallax is listed as $\pi_{\rm RAVE}$ = 4.51 $\pm$ 0.8\,mas which corresponds to a distance of 222 $^{+50}_{-35}$\,pc, while Gaia DR1 tabulates $\pi_{\rm Gaia}$ = 8.72 $\pm$ 0.53\,mas, i.e.\ implying a distance only half as large as the one derived by us. The distance discrepancy is thus confirmed. If interpreted as caused by a partial Dyson sphere, this would convert into $f_\mathrm{cov}\approx 0.77$ and a dimming of the star by $\approx 1.6$ mag at optical/near-IR wavelength. 

What can cause this $\approx$100\% distance offset between Gaia DR1 and the spectroscopic analyses? In practical terms, the spectrophotometric parallax ($\log \pi_{\rm spec} = 0.5([g] - [M]) - 2[{T_\mathrm{eff}}] -0.2(V - A_V + {\rm BC}_V +0.25)$ with [$X$]:=$\log$($X/X_\odot$)) reacts most sensitively to changes in $\log g$, as this parameter has the largest relative uncertainty in the logarithm. However, an unrealistically large $\log g$ correction of $\approx$0.6\,dex (a factor of 4 in $g$) would have to be applied to align the spectrophotometric parallax with the astrometric one. By tweaking the spectroscopic analysis the discrepancy can thus be diminished by a few tens of percent only. 

Additional clues come from radial-velocity follow-up we have conducted since February 2017. The FIES spectrum points to a heliocentric radial velocity of $-4.66$ $\pm$ 0.5\,km/s, fully compatible with the value published by RAVE ($-4.728$ $\pm$ 1.097\,km/s). However, a Magellan/MIKE spectrum taken on 2017-08-13 gives $-25.3$ $\pm$ 1.0\,km/s instead. More recently, two measurements with Mercator/HERMES taken six days apart show a return to values closer to zero, but also indicate short-term variations: $-5.40$ $\pm$ 0.05\,km/s (2018-03-17) and $-5.89$ $\pm$ 0.05\,km/s (2018-03-23). Clearly, TYC 6111-1162-1 is a single-lined spectroscopic binary (SB1). We don't have nearly enough data to constrain the binary-system parameters, and thus only make order-of-magnitude calculations to check whether the binary nature of this system {\em by itself\/} can explain the discrepant distance information. 

To approximately double the parallax from 4.2\,mas (our best estimate) to 8.7\,mas (Gaia's best DR1 estimate) requires an additional $\approx$4\,mas wobble that is not caused by the distance-related trigonometric parallax. This assumes that the spectroscopic analysis of the SB1 spectrum is not affected by the binary companion. At 250\,pc, 4\,mas corresponds to 1\,AU, that is, the secondary has to move the primary by this distance as the two stars circle their common center of mass. Assuming circular orbits, this can be achieved by a $\approx$1\,M$_\odot$ star at a distance of $\approx$1\,AU. This star cannot be a main-sequence star, as it would contribute $\approx$25\% of the photons at optical wavelengths which would surely be visible in the spectrum. Instead, it has to be a white dwarf at the high-mass end of the white-dwarf mass distribution. This scenario thus explains a) the measured radial-velocity amplitude (FIES vs MIKE) and b) the difference between spectrophotometric and trigonometric distance (FIES vs Gaia). It is more or less compatible with the short-term radial-velocity variation which with an orbital velocity of the primary of 20\,km/s (lower limit) would be 0.33\,km/s (0.49\,km/s as measured by HERMES). All of these are modulo orbital ellipticities and projection effects due to the unknown inclination of the orbital plane of the system with respect to the line of sight.

While this scenario is astrophysically plausible and attractive for explaining an extreme outlier as an extreme binary system, it does not explain why and how the repeat observations compiled in Gaia DR1 (14 months of data with a few dozen individual measurements) resulted in a consistent answer $\pi_{\rm Gaia}$ = 8.72 $\pm$ 0.53\,mas in the presence of a time-variable error source of order 4\,mas. Future Gaia data releases (DR2 still treats stars as single objects, while DR3 lifts this assumption) will likely shed light on this question\footnote{Note added in proof: Gaia DR2, which was released while this paper was in review, confirms that the Gaia DR1 parallax is flawed and claims a significantly smaller parallax of $\pi_{Gaia}=5.75\pm 0.17$. However, this DR2 parallax is still based on the assumption of a single-star
solution. We therefore expect the agreement with our spectrophotometric estimate to become even better with Gaia DR3.}. 

\section{Discussion}
\label{discussion}
While the number of stars for which errors on spectrophotometric and parallax distances are sufficiently small to allow Dyson-sphere candidates to be reliably identified remains relatively small at the current time ($\sim 10^4$ objects), the sample is expected to increase significantly in the coming years. Gaia DR2 will provide parallaxes for 1.3 billion objects (compared to 2 million for DR1) which will immediately increase the overlap with ground-based surveys like RAVE and GALAH and thereby boost the number of objects for which the technique can be applied, albeit likely only by factors of a few. Gaia DR2 will also feature luminosity and extinction estimates for $\approx 7.7\times 10^7$ objects. However, since the stellar parameters will be based on photometry and not on spectroscopy, strong degeneracies between effective temperature and dust reddening are expected, and it is not clear that this will increase the number of useful objects for our purposes. By Gaia DR3\footnote{scheduled for late 2020}, these degeneracies will be significantly reduced through the use of data from the Gaia Radial Velocity Spectrometer (RVS) data, and based on the projected performance \citep{Recio_Blanco16,Dafonte16}, we estimate that $\sim 10^6$ stars will have sufficiently small spectrophotometric distance errors to be included in searches like ours. If we assume the same errors as in the cuts used in the current paper, a sample of $\sim 10^6$ objects could give $\sim 10^3$ spurious detections at $f_\mathrm{cov}>0.7$. However, multi-epoch photometry (Gaia and Pan-starrs) combined with archival multi-wavelength data can likely prune the sample further, so that only a minor fraction of these need to be targeted by spectroscopic observing campaigns. Additionally, Gaia DR3 will identify non-single stars, thereby making it easier to weed out binaries like TYC 6111-1162-1.  

We can, however, foresee a couple of complications for our Dyson-sphere search that are likely to linger despite upcoming improvements in Gaia data, which we discuss in the following.

\subsection{The grey-absorber assumption}
One important assumption in the proposed method is that a Dyson sphere would block a fraction $f_\mathrm{cov}$ of the stellar light at optical/near-IR wavelengths, but otherwise not distort the shape of the spectrum in this wavelength range. A violation of this assumption (for instance through light reflected off Dyson-sphere elements into our line of sight, or some non-thermal emission process) would affect the model fit in unpredictable ways, and could, for instance, make bona-fide Dyson spheres in a sample of stars end up in the part of the distribution with very poor spectroscopic fits. Because such objects would supposedly be assigned very large distance errors, our current approach would bias against them, since we have argued that high-$f_\mathrm{cov}$ candidates can only be singled out (using the technique outlined in this paper) from the subset of objects for which distance errors are small.

\subsection{Grey dust}
One effect that potentially could give rise to spurious high-$f_\mathrm{cov}$ Dyson-sphere candidates is grey dust, i.e.\ line-of-sight material that obscures starlight without causing any significant reddening of the spectrum. Any dust component with large, micrometer-sized grains would effectively be grey at optical/near-IR wavelengths, and could -- at sufficiently high optical depth -- reduce the apparent brightness of a star in a way that would be very similar to the expected signatures of a Dyson sphere at these wavelengths. Such large dust grains have been seen forming in the circumstellar material around supernovae \citep{Gall14,Nielsen18} and have also been reported in the interstellar medium \citep[e.g.][]{Wang15}. High levels of grey extinction have also been claimed along certain sight lines \citep[][]{Gorbikov10}, occasionally amounting to several magnitudes of optical extinction \citep[e.g.][]{Skorzynski03,Krelowski16}. However, it has been argued that grey extinction correlates with the strengths of certain diffuse interstellar bands in the optical region \citep[e.g.][]{Skorzynski03}. High signal-to-noise spectroscopy of the spectral regions featuring these bands may therefore be a suitable way to further scrutinize potential Dyson-sphere candidates. 

\section{Summary}
\label{summary}
We have presented a method that, through a comparison of trigonometric and spectrophotometric parallaxes/distances, makes it possible to single out candidates for nearly complete Dyson spheres, even in cases where the Dyson-sphere IR-excess signature is much smaller than usually assumed. The Gaia mission will eventually provide both of the required distances (with highly variable accuracy) and we estimate that Gaia DR 3 (currently scheduled for late 2020) should allow the technique to be applied to samples of $\sim$$10^6$ stars. A small-scale version of this search can also be implemented at the current time by combining Gaia parallaxes with spectrophotometric distances from a ground-based survey. By combining parallax data from Gaia DR1 with spectrophotometric distances from the RAVE survey, we find that $\approx 8000$ stars ($\approx$ 4\% of the objects in common between both surveys) have sufficient data quality to allow potential Dyson-sphere candidates to be efficiently singled out. One of the candidates identified this way is TYC-6111-1162-1, an F-type star with an apparent factor-of-2 distance discrepancy (as would be expected for a Dyson sphere with covering fraction $f_\mathrm{cov}\approx 0.75$), but no detectable mid-IR excess. Through new observations of TYC-6111-1162-1 with the NOT/FIES high-resolution spectrograph, we largely confirm the stellar parameters found by RAVE (with the exception of the metallicity), i.e.\ the star remains a marked outlier. Additional observations with the Magellan/MIKE high-resolution spectrograph reveal significant radial-velocity differences ($\approx 20$\,km/s) with respect to the RAVE and NOT/FIES observations. This indicates that TYC-6111-1162-1 is part of a binary system with an unseen companion. We find it plausible that the astrometric wobble resulting from this fact has interfered with the Gaia DR1 astrometric solution and that, for this system, the astrometric parallax has been overestimated. Taken at face value (given the poor sampling of the radial-velocity curve, assuming circular orbits and ignoring unknown projection effects), a white dwarf of $\approx$1\,M$_\odot$ is needed to double the parallax. Later Gaia data releases may shed further clues on this scenario. Overall, weeding out this type of interloper should become easier with Gaia DR3 which will feature astrometric solutions for binary stars.   

\acknowledgments
\section*{Acknowledgments}
E.Z. acknowledges funding from the Magnus Bergvall foundation and Nordenskj\"{o}ldska Swendeborgsfonden. 
A.J.K. acknowledges long-term support by the Swedish National Space Board for work as part of the {\it Gaia} Data Processing and Analysis Consortium (DPAC). 
This work is based on observations made with the Nordic Optical Telescope, operated by the Nordic Optical Telescope Scientific Association at the Observatorio del Roque de los Muchachos, La Palma, Spain, of the Instituto de Astrofisica de Canarias. \\
We would like to thank Alex Ji, Anna Frebel, Peter de Cat, Sophie van Eck and Alain Jorissen for providing additional radial-velocity measurements.\\ 
This work has made use of data from the European Space Agency (ESA) mission {\it Gaia} (\url{https://www.cosmos.esa.int/gaia}), processed by
the {\it Gaia} DPAC (\url{https://www.cosmos.esa.int/web/gaia/dpac/consortium}). Funding for the DPAC has been provided by national institutions, in particular the institutions participating in the {\it Gaia} Multilateral Agreement.\\
Funding for RAVE has been provided by: the Australian Astronomical Observatory; the Leibniz-Institut fuer Astrophysik Potsdam (AIP); the Australian National University; the Australian Research Council; the French National Research Agency; the German Research Foundation (SPP 1177 and SFB 881); the European Research Council (ERC-StG 240271 Galactica); the Istituto Nazionale di Astrofisica at Padova; The Johns Hopkins University; the National Science Foundation of the USA (AST-0908326); the W. M. Keck foundation; the Macquarie University; the Netherlands Research School for Astronomy; the Natural Sciences and Engineering Research Council of Canada; the Slovenian Research Agency; the Swiss National Science Foundation; the Science \& Technology Facilities Council of the UK; Opticon; Strasbourg Observatory; and the Universities of Groningen, Heidelberg and Sydney.
The RAVE web site is at \url{https://www.rave-survey.org}.

\bibliography{refs}




\end{document}